\documentclass{ifacconf}
\usepackage{amsmath}
\usepackage{amssymb}
\usepackage{mathtools}
\usepackage{algorithm}
\usepackage[noend]{algpseudocode}
\usepackage{graphicx}      
\usepackage{natbib}        

\makeatletter
\def\BState{\State\hskip-\ALG@thistlm}
\makeatother

\DeclarePairedDelimiter\abs{\lvert}{\rvert}%

\usepackage{tikz}
\newcommand\copyrighttext{%
  \footnotesize Accepted to IFAC Workshop on Control of Smart Grid and Renewable Energy Systems (CSGRES), June 10-12, 2019, Jeju, Korea. \\
  \textcopyright  \ 2019 IFAC. This manuscript version is made available under the CC-BY-NC-ND 4.0 license: https://creativecommons.org/licenses/by-nc-nd/4.0/}
\newcommand\copyrightnotice{%
\begin{tikzpicture}[remember picture,overlay]
\node[anchor=south,yshift=53pt] at (current page.south) {\fbox{\parbox{\dimexpr\textwidth-\fboxsep-\fboxrule\relax}{\copyrighttext}}};
\end{tikzpicture}%
}

\begin{document}
\begin{frontmatter}

\title{Estimation of the Shapley Value of a Peer-to-Peer Energy Sharing Game using Coalitional Stratified Random Sampling
\thanksref{footnoteinfo}} 

\thanks[footnoteinfo]{This work was supported in part by the Engineering and 
Physical Sciences Research Council under Grants EP/N03466X/1 and 
EP/S000887/1, and in part by the Oxford Martin Programme on Integrating 
Renewable Energy.}

\author[First]{Liyang Han} 
\author[First]{Thomas Morstyn} 
\author[First]{Malcolm McCulloch}

\address[First]{Department of Engineering Science, University of Oxford, UK 
(e-mail: liyang.han, thomas.morstyn, malcolm.mcculloch@eng.ox.ac.uk).}

\begin{abstract}                
Various peer-to-peer energy markets have emerged in recent years in an attempt to manage distributed energy resources in a more efficient way. One of the main challenges these models face is how to create and allocate incentives to participants. Cooperative game theory offers a methodology to financially reward prosumers based on their contributions made to the local energy coalition using the Shapley value, but its high computational complexity limits the size of the game. This paper explores a stratified sampling method proposed in existing literature for Shapley value estimation, and modifies the method for a peer-to-peer cooperative game to improve its scalability. Finally, selected case studies verify the effectiveness of the proposed coalitional stratified random sampling method and demonstrate results from large games.
\end{abstract}


\end{frontmatter}

\copyrightnotice

\section{Introduction}

The increasing penetration of distributed energy resources (DER) poses challenges to distribution network operation. One of the most important topics recent researches have been focusing on is how to maintain the reliability of energy supply while encouraging distributed renewable generation, which is highly variable and intermittent, see \cite{J.SkeaD.AndersonT.GreenR.Gross2007}. Curtailment is applied in some networks with high renewable generation, see \cite{KLINGEJACOBSEN2012663}. However, it introduces inefficiency into the energy system and financially penalizes owners of renewable resources. Centralized control of DER is also proposed in various researches, but they tend to overlook the fact that prosumers, proactive-consumers with distributed energy resources that actively control their energy behaviors, are independent entities who need incentives to participate in such a centralized control scheme, see \cite{Morstyn2017}. The idea of setting up a peer-to-peer (P2P) energy sharing scheme is gaining tremendous attention both in the industry and academia in recent years, as it is considered a key market strategy to financially encourage efficient local management of DER, see \cite{Parag2016}.

A key feature of a P2P sharing scheme is its ability to use local flexibility to offset generation uncertainty. Local flexibility often takes the form of energy storage (ES), which can be modeled in a similar way as other types of flexible demand, see \cite{7457676}. This paper makes the common assumption that the price to export energy to the energy network is lower than the price to import, see \cite{Zhou2018}; hence, for a single prosumer, the benefit of flexibility is easily reflected in their energy bills when they increase the local usage of their own generation by optimally scheduling their ES. In a P2P market, more joint benefit can be reaped from matching local flexibility with variable generation among all the participants. At the same time, however, it becomes a challenge to allocate the benefit to each participant in an efficient and fair way.

Game theory has been adopted in some recent research to look at how to affect prosumer behavior using financial incentives. Dynamic pricing coupled with non-cooperative game theory is one of the most popular topics, see \cite{LJia2016}, but it fails to demonstrate consistent benefit for every participant, see \cite{8417894}. Cooperative game theory is proposed as an alternative approach and is shown in \cite{8443054} that financial rewards can be fairly allocated using the Shapley value, which is based on the contribution each prosumer makes to this joint scheme. 

A player's Shapley value in a cooperative game is a weighted average of their marginal contribution to all the possible coalitions among all players, see \cite{Shapley1971}. For an \(N\)-player game, there are \(2^N\) possible coalitions, which means that the computation of Shapley value becomes intractable when increasing the number of players. The estimation of the Shapley value has been explored in some previous literature, sampling being the main methodology. An example is the cooperative scheme described in \cite{Chapman2017}, but the model is constructed as simple games with a binary outcome representing whether a coalition of battery-owning households can overcome a hard network constraint. This scheme overlooks the contributions made by local generation or ES units that are not big enough to switch the binary outcome. 

To improve the scalability of the P2P cooperative game proposed in \cite{8443054}, this paper identifies a random sampling method as a way to estimate the Shapley value. The method was proposed in \cite{Castro2009}, and then modified in \cite{Castro2017} by adding a stratification step to the sampling, improving the accuracy of the estimation. This paper adapts this stratified sampling method by further creating \emph{coalitional strata} for better performance in this specific application. The proposed sampling method enables the P2P game to scale up, and we are then able to analyze the impact of different DER adoption rates on prosumer profitability. Some interesting findings are shown in the case studies. 



\section{P2P Cooperative Game}

In an $N$-player cooperative game, the \textit{grand coalition} $\mathcal{N}$ is defined as the group of all $N$ players. Any subset of the \textit{grand coalition} $\mathcal{T}: \mathcal{T} \subseteq \mathcal{N}$ is called a \textit{coalition}. The basic framework of the P2P cooperative game proposed in \cite{8443054} involves mainly three steps. Step 1 is to cooperatively manage DER within all coalitions, which requires optimally scheduling the ES units to minimize the \textit{coalitional energy cost}, see Subsection \ref{subs_coa_e_manag}. Step 2 is to quantify the value of forming each coalition, see Subsection \ref{subs_coa_v}. Step 3 is to divide the total energy cost savings from forming the \textit{grand coalition} to all the players based on certain criteria, see Subsection \ref{subs_pros_pay}.

\subsection{Coalitional Energy Management} \label{subs_coa_e_manag}

We index each prosumer by $i$ and the \textit{grand coalition} by \(i \in \mathcal{N} := \{1,2,...,N\}\). If we consider \(K\) timesteps (\(t = 1, 2,..., K\)) with a time interval of \(\Delta t\), the total energy cost of a coalition $\mathcal{T}$ can be written as
\begin{equation*}
F_{\mathcal{T}}(\mathbf{b}) = \sum_{t=1}^{K} \sum_{i \in \mathcal{T}} \Big\{r^{im}_{t}  [p_{it} + b_{it}]^+ + r^{ex}_{t} [p_{it} + b_{it}]^- \Big\}
\end{equation*}
where \emph{subscripts} $i$ and $t$ are indices for the player and the timestep respectively. The known inputs are $r^{im}_{t}$, $r^{ex}_t$, and $p_{it}$, which are electricity import price (\pounds/\textit{kWh}), electricity export price (\pounds/\textit{kWh}), and net energy consumption (positive) or generation (negative) (\textit{kWh}) without ES. The variables are $\mathbf{b}  \in \mathbb{R}^{N \times K} := b_{it}, \forall i \in [1,N], \forall t \in [1,K]$: ES charge (positive) or discharge (negative) energy (\textit{kWh}). We also define operation \([z]^{+(-)} = \max (\min) \{z,0\}\).

With the assumption $r^{im}_t > r^{ex}_t, \forall t$, we can schedule all the ES units' operation within coalition $\mathcal{T}$ to minimize the \emph{coalitional energy cost} $G(\mathcal{T})$, which is defined as
\begin{align}
G(\mathcal{T}) = \ & \min_{\mathbf{b}} {F_\mathcal{T}(\mathbf{b})} \nonumber \\
s.t. \quad & \underline{b}_{i} \leq b_{it} \leq \overline{b}_{i}, \ \ \forall i \in \mathcal{T}, \forall t \in [1, K] \label{pconst} \\
& 0 \leq e_{i} SoC^{0}_i + \sum_{t=1}^k ([b_{it}]^{+} \eta_{i}^{in} + [b_{it}]^{-} / \eta_{i}^{out})  \leq e_{i} \nonumber \\
& \qquad \qquad \qquad \qquad \qquad \quad \forall i \in \mathcal{T}, \forall k \in [1, K] \label{econst} \\
& \sum_{t=1}^K ([b_{it}]^{+} \eta_{i}^{in} + [b_{it}]^{-} / \eta_{i}^{out}) = 0, \ \ \forall i \in \mathcal{T} \label{cconst}
\end{align}
where (\ref{pconst}), (\ref{econst}), and (\ref{cconst}) represent the ES power constraint, energy constraint, and cycle constraint respectively. We consider each prosumer \(i\)'s ES system has an \textit{energy capacity} \((\mathit{kWh})\) of \(e_{i} \geq 0\), a \textit{charge limit} \((\mathit{kWh})\) of \(\overline{b}_{i} \geq 0\) and a \textit{discharge limit} \((\mathit{kWh})\) of \(\underline{b}_{i} \geq 0\) over the time span of \(\Delta t\), a \textit{charge efficiency} of \(\eta_{i}^{in} \in (0, 1)\) and a \textit{discharge efficiency} of \(\eta_{i}^{out} \in (0, 1)\), and an \textit{initial state of charge} of \(SoC^{0}_i \in [0, 1]\). For a prosumer who does not own an ES system, we set their energy capacity and charge/discharge limits all as zeros. 

\begin{figure}
\begin{center}
\includegraphics[width=8.4cm]{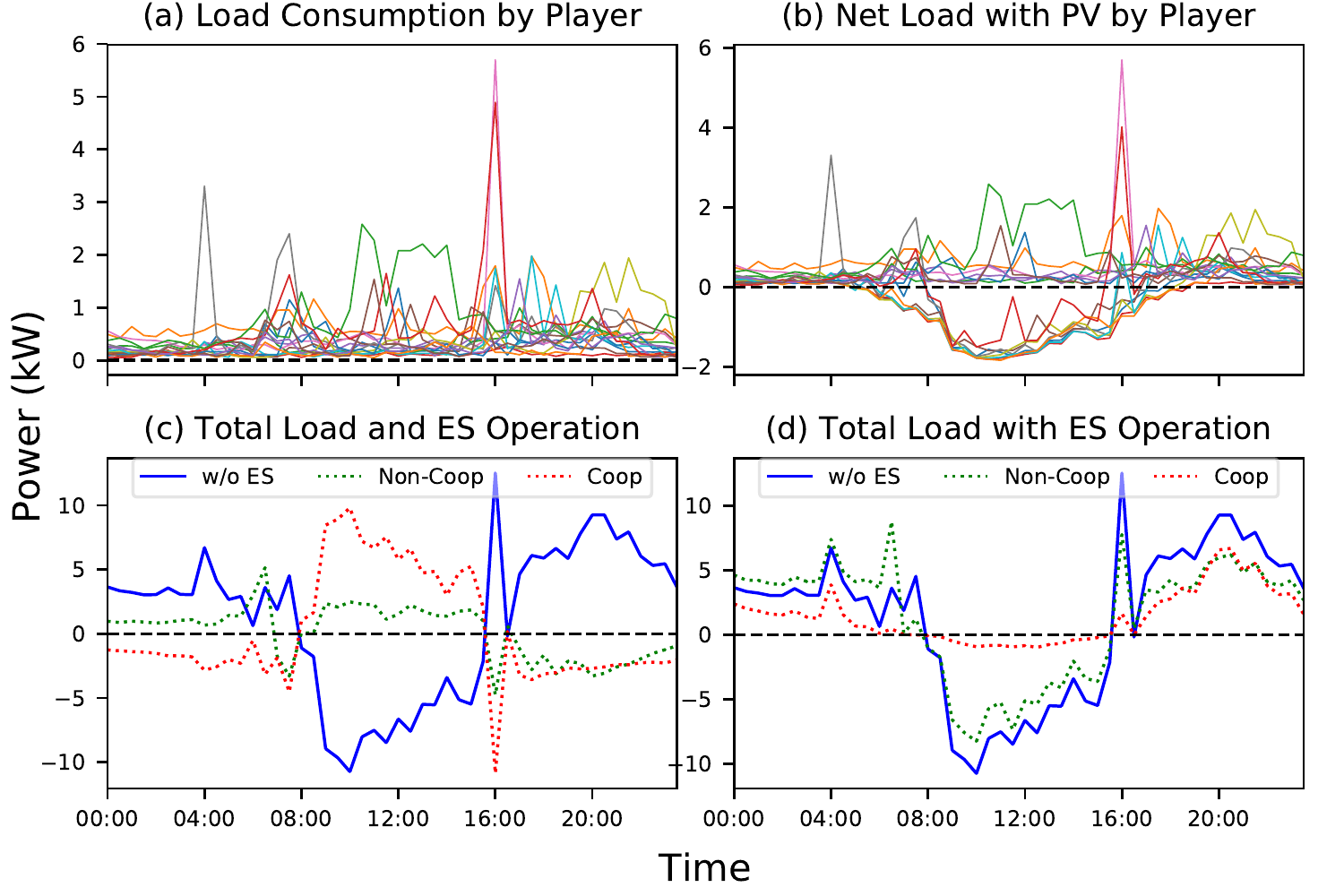}    
\caption{16 Prosumer loads: (a) individual load consumption, (b) individual load including PV generation, (c) grand coalition load and non-cooperative and cooperative ES operation profiles, (d) grand coalition load with non-cooperative and cooperative ES operation.} 
\label{fig:loads_sum}
\end{center}
\end{figure}

Fig. \ref{fig:loads_sum} demonstrates the effect of cooperative ES operation in a 16-prosumer scenario. As shown in (d), the cooperative ES operation tends to flatten the load as it tries to match the consumption and generation within the coalition to minimize the \emph{coalition energy cost}.

\subsection{Value of Coalitions} \label{subs_coa_v}

The purpose of using cooperative game theory is to establish a framework to quantify the benefit of cooperation, and then to allocate the benefit to the participants efficiently. The \emph{coalitional energy cost} provides a great metric to evaluate a coalition's performance. Here, we define the value of a coalition  \(\mathcal{T}\) as the energy cost savings obtained by forming the coalition. This is given by the difference between the sum of the energy costs incurred by each prosumer in \(\mathcal{T}\) when they schedule the ES systems individually, and the minimum \textit{coalitional energy cost} of \(\mathcal{T}\) when the prosumers schedule their ES systems collectively:
\begin{equation*}
v(\mathcal{T}) = \sum_{i \in \mathcal{T}} {G({\{i\}})} - G({\mathcal{T}})
\end{equation*}

By this definition, the value of the \emph{grand coalition} becomes the total energy cost savings of a P2P cooperative game, which denotes the total amount of payoffs we can award to all the participants.

\subsection{Prosumer Payoffs and Shapley Value} \label{subs_pros_pay}

The second step in a cooperative game framework is the allocation of payoffs. We use vector \(\mathbf{x} \in \mathbb{R}^{N}\) as the \textit{payoff allocation} whose entry \(x_i\) represents the payment to prosumer \(i \in \mathcal{N}\). One important payoff allocation is called the Shapley value denoted as $\phi_i, \forall i \in \mathcal{N}$, see \cite{Shapley1971}, representing each player's weighted average marginal contribution to all possible coalitions within the game:
\begin{equation}
\phi_i = \sum_{\mathclap{{\mathcal{T}  \subset \mathcal{N}, i \notin \mathcal{T}}}} \frac{(\abs{\mathcal{T}})! (N - \abs{\mathcal{T}} - 1)!}{N!} [v(\mathcal{T} \cup i) - v(\mathcal{T})]
\label{shapley_calc}
\end{equation}

The Shapley value also satisfies the following axioms:
\begin{enumerate}
\item (\textit{Efficiency}) \(\sum_{i \in \mathcal{N}} {\phi_i} = v(\mathcal{N})\). This requires the entirety of the value created by the \textit{grand coalition} to be allocated to the players. \label{axm_eff}
\item (\textit{Individual Rationality}) \(\phi_i \geq v(\{i\}), \forall i \in \mathcal{N}\). This ensures that no player is penalized for cooperating.  \label{axm_ind}
\item (\textit{Symmetry}) If \(v(\mathcal{T} \cup \{i\}) = v(\mathcal{T} \cup \{j\}), \forall \mathcal{T} \subseteq \mathcal{N}, \mathcal{T} \cap \{i,j\} = \emptyset\), then \(\phi_i = \phi_j \).  This means that two players should be assigned the same Shapley value if they have the same marginal contributions to all the coalitions. \label{axm_sym}
\item (\textit{Dummy Axiom}) If \(v(\mathcal{T}) = v(\mathcal{T} \cup \{i\}), \forall \mathcal{T} \subseteq \mathcal{N}, \mathcal{T} \cap \{i\} = \emptyset\), then \(\phi_i = 0\). Therefore, a player's Shapley value should be zero if they add zero marginal value to any of the coalitions. \label{axm_dum}
\item (\textit{Additivity}) If \(v\) and \(u\) are \text{characteristic functions}, then \(\phi_i(v+u) = \phi_i(v) + \phi_i(u), \forall i \in \mathcal{N}\). This indicates that the Shapley value of two games played at the same time should be the sum of the two games' Shapley values when played separately. \label{axm_add}
\end{enumerate}

Axiom (\ref{axm_eff}) guarantees that all the profits allocated to the prosumers add up to the total energy cost savings from the \emph{grand coalition}. In our P2P cooperative game, \(v(\{i\}) = 0, \forall i \in \mathcal{N}\), so Axiom (\ref{axm_ind}) requires \(\phi_i \geq 0, \forall i \in \mathcal{N} \). Axiom (\ref{axm_sym}) and (\ref{axm_dum}) ensure the \emph{`fairness'} of the payoff allocation. Axiom (\ref{axm_add}) is not actively used in this paper as the P2P cooperative game is the only game discussed here.

The Shapley value offers a way to incentivize prosumers to participate in this cooperative scheme, improving the local energy supply reliability while encouraging the efficient use of distributed renewable generation. However, the scalability of the proposed model is very limited because the Shapley value's computational time increases exponentially with the size of the grand coalition. The following section looks into a sampling method to estimate the Shapley value to reduce the model's computational complexity. 

\section{Estimation of Shapley Value}

The scalability of the P2P cooperative game model is mainly limited by the sheer number of cost minimization problems that are required to be solved. This number is equal to the number of possible coalitions, $2^N$, where $N$ is the number of participating prosumers. Since the Shapley value is the weighted average of a player's marginal contributions, sampling is identified as a promising estimation technique to be applied in our P2P cooperative game.








\subsection{Stratified Random Sampling}

The conventional definition is expressed in (\ref{shapley_calc}). \cite{RePEc:cwl:cwldpp:471r} provided an alternative definition of the Shapley value expressed in terms of all possible orders of the players, which was then adopted by \cite{Castro2009} to develop a random sampling method to estimate the Shapley value. In this approach, $\pi(\mathcal{N})$ is defined as the set of all possible permutations with player set $\mathcal{N}$, and $O: \mathcal{N} \rightarrow \mathcal{N}$ as a permutation that assigns player $O(k)$ to position $k$. For a given $O \in \pi(\mathcal{N})$, the set of predecessors of the player $i$ is denoted as $Pre^i(O)$, where if $i = O(k)$, $Pre^i(O) = \{O(1), O(2), ..., O(k-1)\}$. Player $i$'s marginal contribution is $\delta(O)_i = v(Pre^i(O) \cup i) - v(Pre^i(O))$. The alternative definition of Shapley value can be written as
\begin{equation} \label{sh_alter}
\phi_i = \sum_{O \in \pi(\mathcal{N})} \frac{1}{N!} \delta(O)_i, i \in \mathcal{N}
\end{equation}

Since $\abs{\pi(\mathcal{N})} = N!$ and $\delta(O)_i$ are equally weighted for all $O \in \pi(\mathcal{N})$ in (\ref{sh_alter}), the Shapley value can be estimated using the unweighted expectation of $\delta(O)_i$ given a set of randomly sampled permutations $M$:
\begin{equation} \label{sh_est_rand}
\phi_i = \sum_{O \in M} \frac{1}{\abs{M}} \delta(O)_i, i \in \mathcal{N}
\end{equation}

The alternative definition of the Shapley value is just a special case where $M=\pi(\mathcal{N})$.

Using (\ref{sh_est_rand}), the Shapley value can be estimated from randomly sampling player permutations, see \cite{Castro2009}. To improve the estimation accuracy, \cite{Castro2017} proposed a stratified random sampling approach to divide the population of all player permutations into subpopulations that have the same size of predecessors for each player. This stratified random sampling method follows the following steps.
\begin{enumerate}
\item A stratum, or a stratified set of player permutations is defined as $P_{il} := \{O \in \pi(\mathcal{N}) \mid O(l) = i, \forall i, l \in [1, N] \}$. Therefore, $P_{il}$ contains every permutation $O \in \pi(\mathcal{N})$, in which player $i$ is in position $l$. Player $i$'s mean marginal contribution of each stratum is
\begin{equation} \label{mar_stra}
\phi_{il} = \frac{1}{\abs{P_{il}}} \sum_{O \in P_{il}} \delta(O)_i, \forall i,l \in [1, N]
\end{equation}
\item A random permutation sample $M_{il}$ of size $\abs{M_{il}}$ is obtained with replacement from each stratum $P_{il}$.
\item {Adapted from (\ref{mar_stra}), player $i$'s mean marginal contribution of the samples from each stratum is
\begin{equation} \label{mar_stra_smpl}
\overline{\phi}_{il} = \frac{1}{\abs{M_{il}}} \sum_{O \in M_{il}} \delta(O)_i, \forall i,l \in [1, N]
\end{equation}
The estimated Shapley value can then be calculated as $\overline{\phi}^{st}_i = \sum_{l=1}^N {\frac{1}{N} \overline{\phi}_{il}}, \forall i \in [1, N]$.
}
\end{enumerate}

We notice that in (\ref{mar_stra_smpl}), $\delta(O)_i$ are equally weighted for all $O \in M_{il}$. Because each player set (coalition) $Pre^i(O)$ appears $(l-1)!(N-l)!$ times for a given $l$, they have the same probability of being sampled from $P_{il}$ into $M_{il}$. We define the \emph{coalitional stratum} as the set of coalitions $Q_{il} := \{\mathcal{T} \subseteq \mathcal{N} \mid i \notin \mathcal{T}, \abs{\mathcal{T}} = l-1, \forall i,l \in [1,N]\}$, and $\Delta(\mathcal{T})_i = v(\mathcal{T} \cup i) - v(\mathcal{T})$. We then obtain a random sample $H_{il}$ with replacement from $Q_{il}$, and because the order of players does not matter in a coalition, $H_{il}$ can be considered a combination sample. (\ref{mar_stra_smpl}) can be rewritten as
\begin{equation} 
\overline{\phi}_{il} = \frac{1}{\abs{H_{il}}} \sum_{\mathcal{T} \in H_{il}} \Delta(\mathcal{T})_i, \forall i,l \in [1, N]
\label{eq:coalition_str}
\end{equation}

\subsection{Modified Sampling with Optimal Sample Allocation}

In order to implement the stratified random sampling method, a procedure to determine the sample size of each stratum needs to be established. \cite{Castro2009} identified the true variance as a metric to allocate the samples among strata to minimize the estimation error, and proposed a two-stage Shapley value estimation algorithm with optimal sample allocation. In the first stage, 50\% of the samples are evenly distributed to each stratum to obtain an initial estimated Shapley value and each stratum's sample variance. In the second stage, the remaining 50\% of the samples are optimally allocated to each stratum in proportion to their sample variances calculated in the first stage. The final estimated Shapley value is then calculated using the sampling results from both stages.



We then recognize that $\abs{Q_{il}} = \frac{(N-1)!}{(l-1)!(N-l)!}$, which means that evenly dividing the samples in the first stage could result in a sample size larger than the size of some \emph{coalitional strata}: $\abs{H_{il}} > \abs{Q_{il}}$, especially when $l$ is close to $1$ or $N$. Applying random sampling to obtain $\overline{\phi}_{il}$ in these \emph{coalitional strata} would take even more time and produce less accurate results than directly calculating $\phi_{il}$:
\begin{equation} 
\phi_{il} = \frac{1}{\abs{Q_{il}}} \sum_{\mathcal{T} \in Q_{il}} \Delta(\mathcal{T})_i, \forall i,l \in [1, N]
\label{eq:real_sh_str}
\end{equation}
Using (\ref{eq:coalition_str}) and (\ref{eq:real_sh_str}), we modify the two-stage stratified random sampling method with optimal sample allocation to estimate the Shapley value. This modified method is detailed in Algorithm \ref{alg_mod_strat}.

\begin{algorithm} 
\caption{Two-Stage Coalitional Stratified Random Sampling with Optimal Sample Allocation} \label{alg_mod_strat}
\begin{algorithmic}[0] 
\State {\textbf{Stage 1}}
\State {$h \gets$ total sample size}
\State {$h^A_{il} \gets \frac{h}{2N^2}$}
\State {$\Omega \gets \varnothing:$ set of strata with sample sizes determined}
\For {$i \in [1,N], l \in [1,N]$}
\If {$h^A_{il} > \abs{Q_{il}} = \frac{(N-1)!}{(l-1)!(N-l)!} $} 
\State {$H^A_{il} \gets Q_{il}, \ h^{tot}_{il} \gets \abs{Q_{il}}$}
\State {$h \gets h - \abs{Q_{il}}, \ \Omega \gets \Omega \cup (i,l)$} 
\Else
\State{$H^A_{il} \gets h^A_{il}$ samples with replacement from $Q_{il}$}
\EndIf
\State {$\Phi_{il} \gets 0, \ s \gets 0$}
\For {$\mathcal{T} \in H^A_{il}$}
\State {$\Phi_{il} \gets \Phi_{il} + \Delta(\mathcal{T})_i, \ s \gets s + (\Delta(\mathcal{T})_i)^2$}
\EndFor
\State $\overline{\sigma}^2_{il} \gets \frac{1}{\abs{H^A_{il}}-1} (s - \frac{(\Phi_{il})^2}{\abs{H^A_{il}}}) $
\EndFor
\hrulefill
\State {\textbf{Stage 2}}
\State {$\omega \gets \{(0,0)\}$: initialize $\omega$ to start following \textbf{while} loop}
\While {$\omega \neq \varnothing$}
\For {$i \in [1,N], l \in [1,N]$ \textbf{and} $(i,l) \notin \Omega$}
\State {$h^{tot}_{il} \gets h \frac{\overline{\sigma}^2_{il}}{\sum_{i=1}^N \sum_{l=1}^N \overline{\sigma}^2_{il}}$}
\State {$h^B_{il} \gets h^{tot}_{il} - h^A_{il}$}
\EndFor
\State {$\omega \gets \varnothing:$ set of over-sampled strata in \textbf{Stage 1}}
\For {$i \in [1,N], l \in [1,N]$ \textbf{and} $(i,l) \notin \Omega$}
\If {$h^B_{il} < 0$}
\State {$h^{tot}_{il} \gets h^A_{il}, \ h \gets h - h^A_{il}, \ \omega \gets \omega \cup (i,l)$}
\EndIf
\EndFor
\State {$\Omega \gets \Omega \cup \omega$}
\EndWhile
\For {$i \in [1,N], l \in [1,N]$ \textbf{and} $(i,l) \notin \Omega$}
\State{$H^B_{il} \gets h^B_{il}$ samples with replacement from $Q_{il}$}
\For {$\mathcal{T} \in H^B_{il}$}
\State {$\Phi_{il} \gets \Phi_{il} + \Delta(\mathcal{T})_i$}
\EndFor
\EndFor
\State {$\overline{\phi}_{il} \gets \frac{\Phi_{il}}{h^{tot}_{il}}$ \textbf{for} $i \in [1,N], l \in [1,N]$}
\State {$\overline{\phi}^{cl,st,opt}_{i} \gets \sum_{l=1}^N \frac{1}{N} \overline{\phi}_{il}$ \textbf{for} $i \in [1,N]$}
\State \Return {$\overline{\phi}^{cl,st,opt}_{i}, i \in [1,N]$}
\end{algorithmic}
\end{algorithm}

In Stage 1, in the case where the evenly distributed sample size is bigger than the stratum size, we compute the stratum's precise mean marginal contribution $\phi_{il}$, and add the saved samples to Stage 2. This way, we can improve the accuracy of the estimation both by using the precise stratum marginal contribution, and by increasing the number of samples for optimal allocation. 

\section{Case Studies}

In the following two case studies, we implement the proposed sampling method to estimate the Shapley value of our P2P cooperative game. In the first case study, we select a range of prosumer numbers so we can compare the computational time of the estimated Shapley value and the actual Shapley value, and evaluate the accuracy of the estimation. In the second case study, we scale up the size of the game to evaluate the payoffs to the prosumers based on their DER types. 

Some of the model inputs are as follows: the domestic load data was measured in the Customer-Led Network Revolution trials\footnote{http://www.networkrevolution.co.uk/resources/project-library}. the model time frame is 24 hours starting from the midnight of a sunny summer day in July. The PV systems are 4kW with fixed 20 degree tilt, simulated in PVWatts\footnote{http://pvwatts.nrel.gov/pvwatts.php} using the London Gatwick solar data. The ES model has an \textit{energy capacity} of 7 kWh, a maximum charge power of 3.5 kW, a maximum discharge power of 3.2 kW, both charge and discharge efficiencies of 95\%, an \textit{initial state of charge} of 50\%, and a \textit{state of charge} range of 20-95\%. The energy import price follows a UK Economy 7 residential rate structure: \pounds 0.072/kWh for midnight--7am, and \pounds 0.1681/kWh for 7am--midnight\footnote{https://www.gov.uk/government/statistical-data-sets/annual-domestic-energy-price-statistics}, and the energy export price is the UK feed-in tariff\footnote{https://www.gov.uk/feed-in-tariffs/overview} fixed at \pounds 0.0485/kWh.

\subsection{Validation of Sampling-Based Shapley Estimation}

In this case study, we fix the the PV and ES adoption rates both at 50\%, and both ownerships are randomly assigned independently of each other. In other words, each prosumer can have a PV system, or an ES system, or both, or neither. We apply a range of prosumer numbers to compare the computation time between the full Shapley value calculation and the Shapley value estimation with the proposed sampling method. 

Compared against each other are three models: 1) full Shapley value calculation, 2) Shapley value estimation using the proposed sampling method with $10^3$ samples per player, and 3) Shapley value estimation using the proposed sampling method with $10^2$ samples per player. 

Table \ref{tb:comp_t} shows the computation time\footnote{Running on Apple iMac with a processor of 2.8 GHz Intel Core i5 and a memory module of 16 GB 1867 MHz DDR3} of the three models. We only show computation times that are under 10 hours as we consider any time above 10 hours to be impractical for this application. As predicted, the full Shapley value calculation is shown to be intractable. When the number of players exceeds 16, the sampling method significantly reduces the computation time, and with the same number of players the computation time is largely in proportion to the number of samples specified.

\begin{table}[htb]
\begin{center}
\caption{Model Computation Time (s)} \label{tb:comp_t}
\vspace*{-0.14cm}
\begin{tabular}{cccccccc}
\hline
\textbf{No. players} & \textbf{8} & \textbf{12} & \textbf{16} & \textbf{20} & \textbf{30} & \textbf{50} \\\hline
full model & 25 & 466 & 1E+4 & N/A & N/A & N/A \\
$10^3$ samples/p & 11 & 187 & 2E+3 & 6E+3 & 2E+4 & N/A \\
$10^2$ samples/p & 10 & 104 & 221 & 741 & 2E+3 & 2E+4 \\\hline
\end{tabular}
\end{center}
\end{table}

\begin{figure}
\begin{center}
\includegraphics[width=8.4cm]{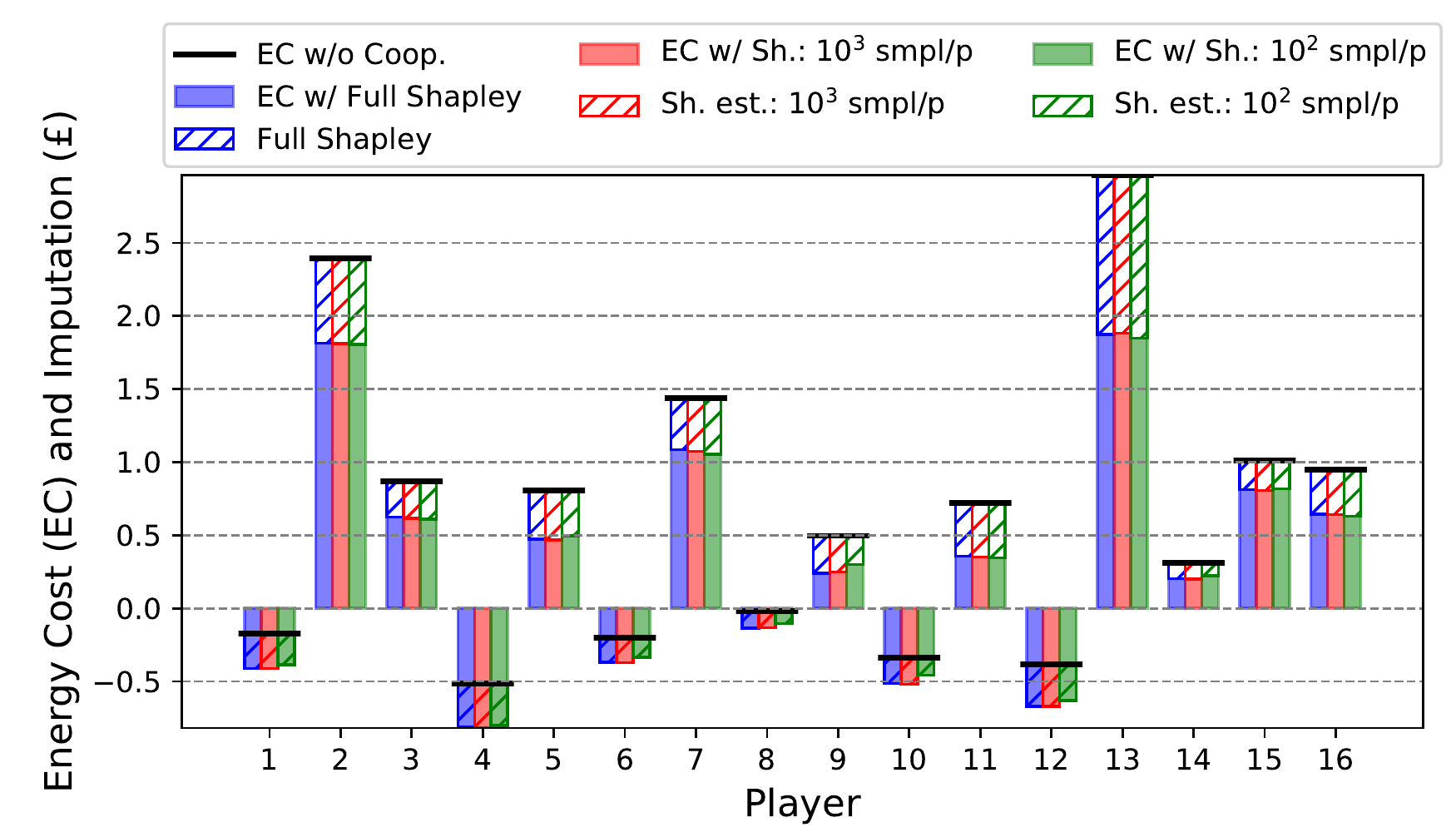}    
\caption{Energy costs (EC) and Shapley values (full model vs. sampling) by player} 
\label{fig:imps_bars}
\end{center}
\end{figure}

We then compare the model results for the 16-player game, the largest game that can be computed for a full model within a reasonable time. From Fig. \ref{fig:imps_bars} we can see that the Shapley value estimation accuracy with $10^3$ samples per player is very high, whereas the estimation with $10^2$ is slightly less accurate. This confirms that there is a trade-off between the computation time and the accuracy of the model when choosing the number of samples. 

\begin{figure}
\begin{center}
\includegraphics[width=7cm]{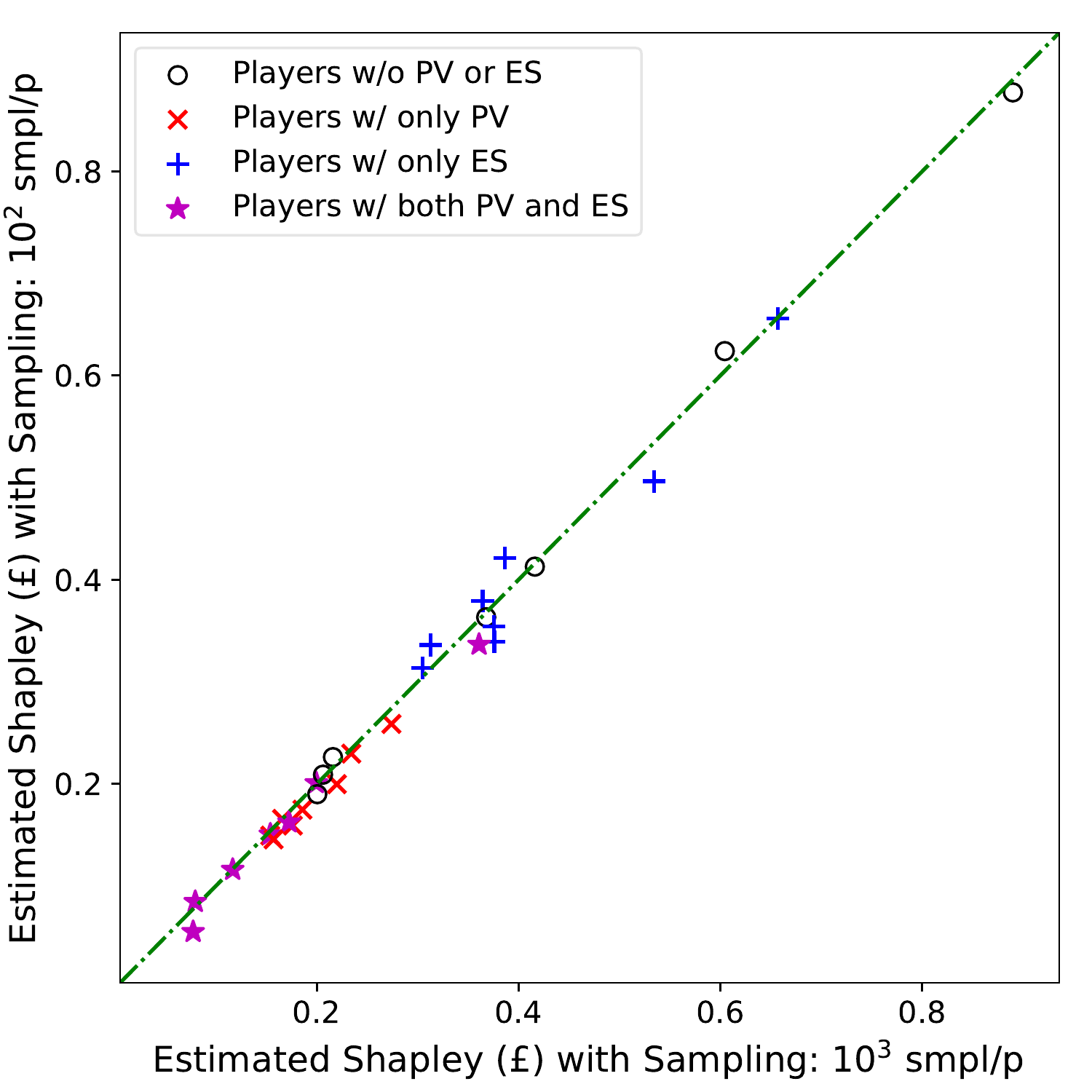}    
\caption{Comparison of the estimated Shapley values using proposed sampling method with different sample sizes} 
\label{fig:est_sh_comp}
\end{center}
\end{figure}

In order to understand this trade-off better when the number of prosumer is further increased. We select a game of 30 prosumers and compare the estimated Shapley values with the two different sample sizes, and the results are plotted in Fig. \ref{fig:est_sh_comp}. Even though the computation time for the $10^3$ samples/player model is about 10 times the $10^2$ samples/player model, see Table \ref{tb:comp_t}, the estimated Shapley values from the two models are very similar regardless of the type of resources owned by a prosumer. This gives us confidence in using a relatively low number of samples ($\geq 10^2$ samples/player) to estimate the Shapley value of larger P2P cooperative games.

\subsection{Sampling-Based Shapley Value for Large Games}\label{SCM}

\begin{figure}
\begin{center}
\includegraphics[width=8.4cm]{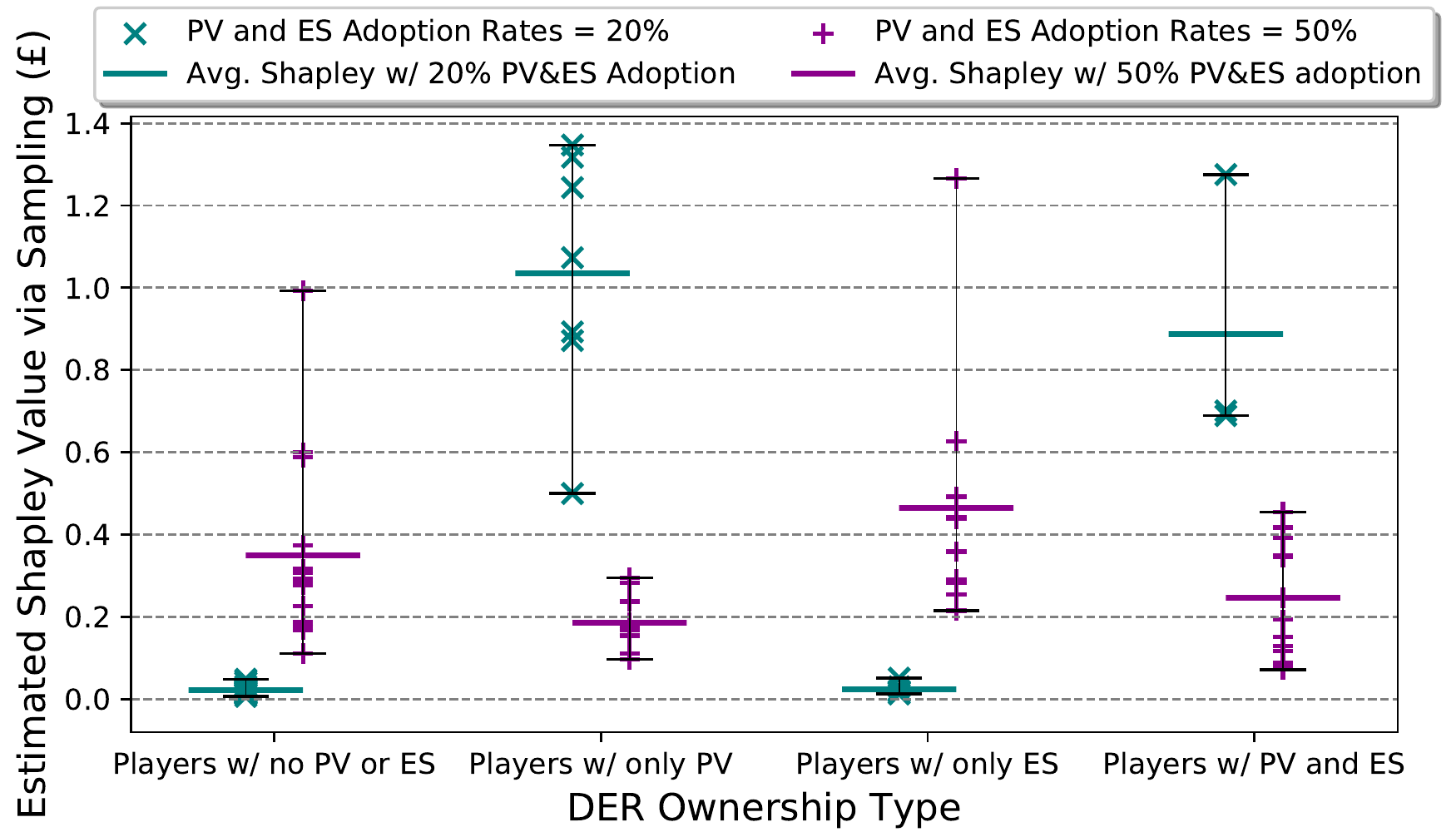}    
\caption{Estimated Shapley value by DER ownership type with two different DER adoption rates (20\% vs. 50\%)} 
\label{fig:est_sh_der}
\end{center}
\end{figure}

For a P2P cooperative game with 50 prosumers, we use 250 samples/player to ensure each \emph{coalitional stratus} is sufficiently represented in the samples while keeping the computation within an acceptable time. First, we keep the number of PV and ES systems the same and vary their adoption rates together. Fig. \ref{fig:est_sh_der} compares the estimated Shapley values by players' DER ownership type, where each marker represents a prosumer's estimated Shapley value. There are a few interesting observations. First, except for a few outliers, prosumers with the same DER ownership type are rewarded similar Shapley values regardless of the overall DER adoption rate. Second, as the DER adoption rates change, there is a significant shift in the Shapley values; when the DER adoption rates are low, PV owners are awarded significantly higher Shapley values likely because they provide cheaper energy to the coalitions, while when the DER adoption rates are high, pure consumers and ES owners are awarded higher Shapley values likely because they absorb more local generation. Third, as the DER adoption rates increase, the average Shapley values by DER ownership type tends to converge despite the wider spread among the pure consumers and prosumers with only ES systems. 

\begin{figure}
\begin{center}
\includegraphics[width=8.4cm]{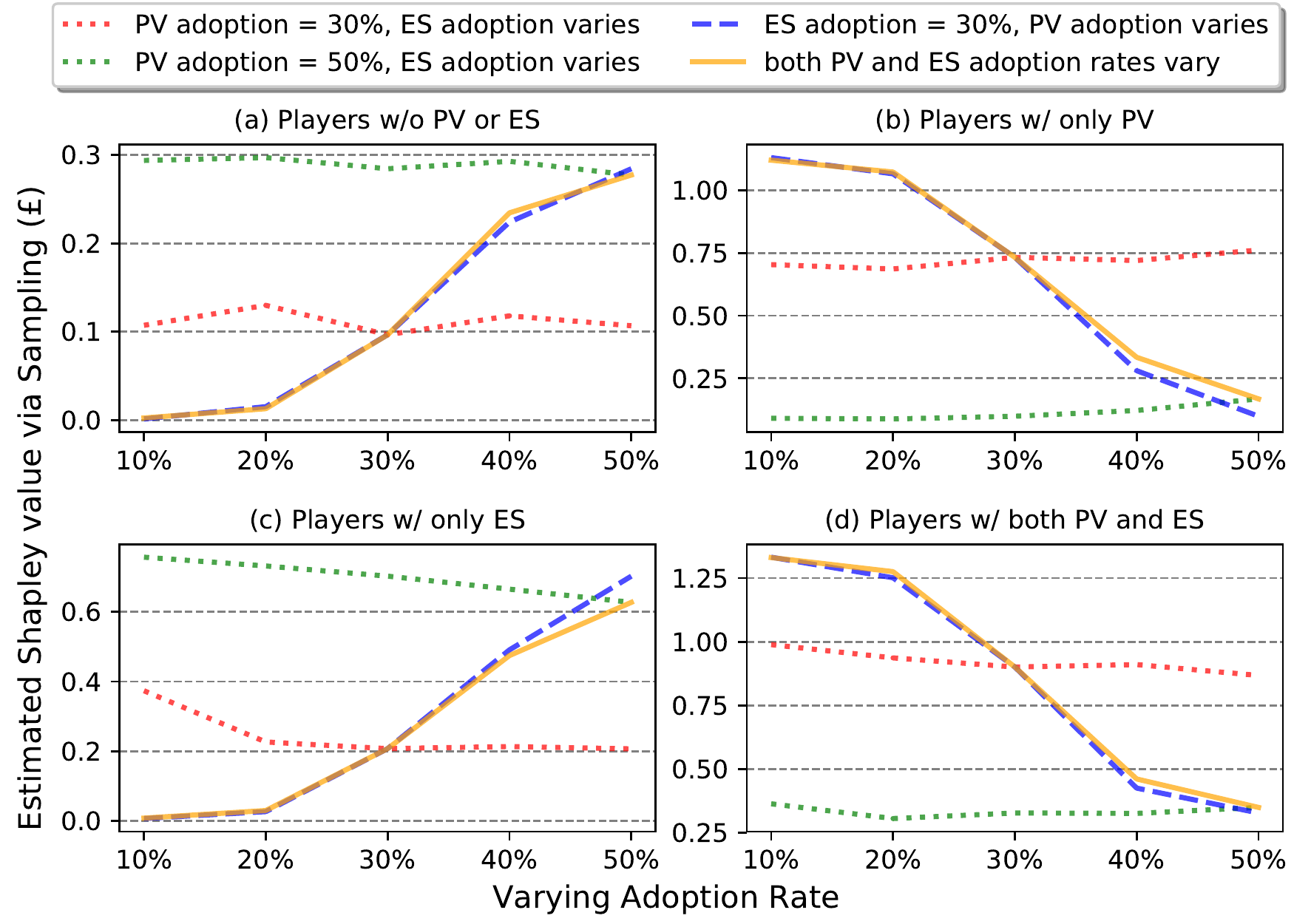}    
\caption{Estimated Shapley value with varying PV and ES adoption rates} 
\label{fig:est_sh_pen}
\end{center}
\end{figure}

With the same 50 prosumers, we then pick out four typical prosumers with different DER ownership types, and run the P2P cooperative model under four different scenarios: 1) PV adoption rate is fixed at 30\%, and ES adoption rate varies from 10\% to 50\%, 2) PV adoption rate is fixed at 50\%, and ES adoption rate varies from 10\% to 50\%, 3) ES adoption is rate fixed at 30\%, and PV adoption rate varies from 10\% to 50\%, and 4) PV and ES are with the same adoption rate that varies from 10\% to 50\%. Fig. \ref{fig:est_sh_pen} illustrates how the Shapley value changes with different DER adoption rates. Based on the DER ownership type, the trend at which the Shapley value changes with the varying DER adoption rates can be very different. For example, a consumer that does not own any PV or ES tends to be awarded more when the adoption rates for the PV and ES increase together, whereas a prosumer that owns both PV and ES display the opposite trend. It is interesting to note that when the PV adoption rate is fixed, whether at 30\% or 50\%, varying the ES adoption rate has very little influence on the Shapley value regardless of the prosumer type. In contrast, whether the ES adoption rate is fixed at 30\% or follows the PV adoption rate, varying the PV adoption rate has a significant impact on the Shapley value of all prosumer types.

It is worth noting that the main purpose of the case studies is to validate the scalability of the proposed sampling method applied in the P2P cooperative game. The specific results shown are dependent on the assumptions made about the PV, ES system specifications, and the energy prices. Further sensitivity analyses need to be conducted to generalize the results to other markets. 

\section{Conclusion}

To improve the scalability of the P2P cooperative game \citep{8443054}, this paper modifies a stratified random sampling method \citep{Castro2017} to estimate the Shapley value. The maximum size of the game that can be computed in a reasonable time ($<10$ hours) is thus increased from less than 20 players to 50 players. Through case studies, the estimation errors are shown to be very small. The proposed model is then run on a P2P cooperative game of 50 players to demonstrate some interesting patterns and trends in the Shapley value for different prosumer DER ownership types and with varying DER adoption rates. Some future work includes sensitivity analyses on the PV and ES system inputs and electricity prices, and improving the sampling method to be able to further scale up the size of the P2P cooperative game.






\bibliography{ifacconf_csgres2019_LH.bib}             
                                                   







\end{document}